%% file: main.tex
\documentclass[10pt,twocolumn,letterpaper]{article}
\usepackage[pagenumbers]{cvpr} 
\definecolor{cvprblue}{rgb}{0.21,0.49,0.74}
\usepackage[pagebackref,breaklinks,colorlinks,allcolors=cvprblue]{hyperref}
\usepackage{booktabs}
\usepackage{graphicx}
\usepackage[normalem]{ulem}
\usepackage[dvipsnames]{xcolor}

\newcommand{\name}{3D-LATTE}
\usepackage[table]{xcolor}

\newcommand{\boldparagraph}[1]{\vspace{0.5em}\noindent{\bf #1.}}

\renewcommand{\paragraph}[1]{\boldparagraph{#1}}

\newcommand{\best}[1]{\cellcolor{red!20}\textbf{{#1}}}
\newcommand{\second}[1]{\cellcolor{orange!20}\uline{{#1}}}
\newcommand{\third}[1]{\cellcolor{yellow!20}#1}

\DeclareRobustCommand{\bestcap}[1]{\colorbox{red!20}{\textbf{#1}}}
\DeclareRobustCommand{\secondcap}[1]{\colorbox{orange!20}{\uline{#1}}}
\DeclareRobustCommand{\thirdcap}[1]{\colorbox{yellow!20}{#1}}

\title{ 3D-LATTE: Latent Space 3D Editing from Textual Instructions}

\author{
Maria Parelli$^{1}$ \qquad
Michael Oechsle$^{2}$ \qquad
Michael Niemeyer$^{2}$ \qquad
 \vspace{4px}
 \\
Federico Tombari$^{2}$ \qquad
Andreas Geiger$^{1}$
\vspace{4px}
\\
\vspace{5px}\\
{ \small
$^{1}$University of Tübingen, Tübingen AI Center\quad
$^{2}$Google Zurich\quad
}}
\begin{document}
\twocolumn[{%
\renewcommand\twocolumn[1][]{#1}%
\maketitle
\begin{center}
    \captionsetup{type=figure}
    \vspace{-0.71cm}
    \includegraphics[width=\textwidth]{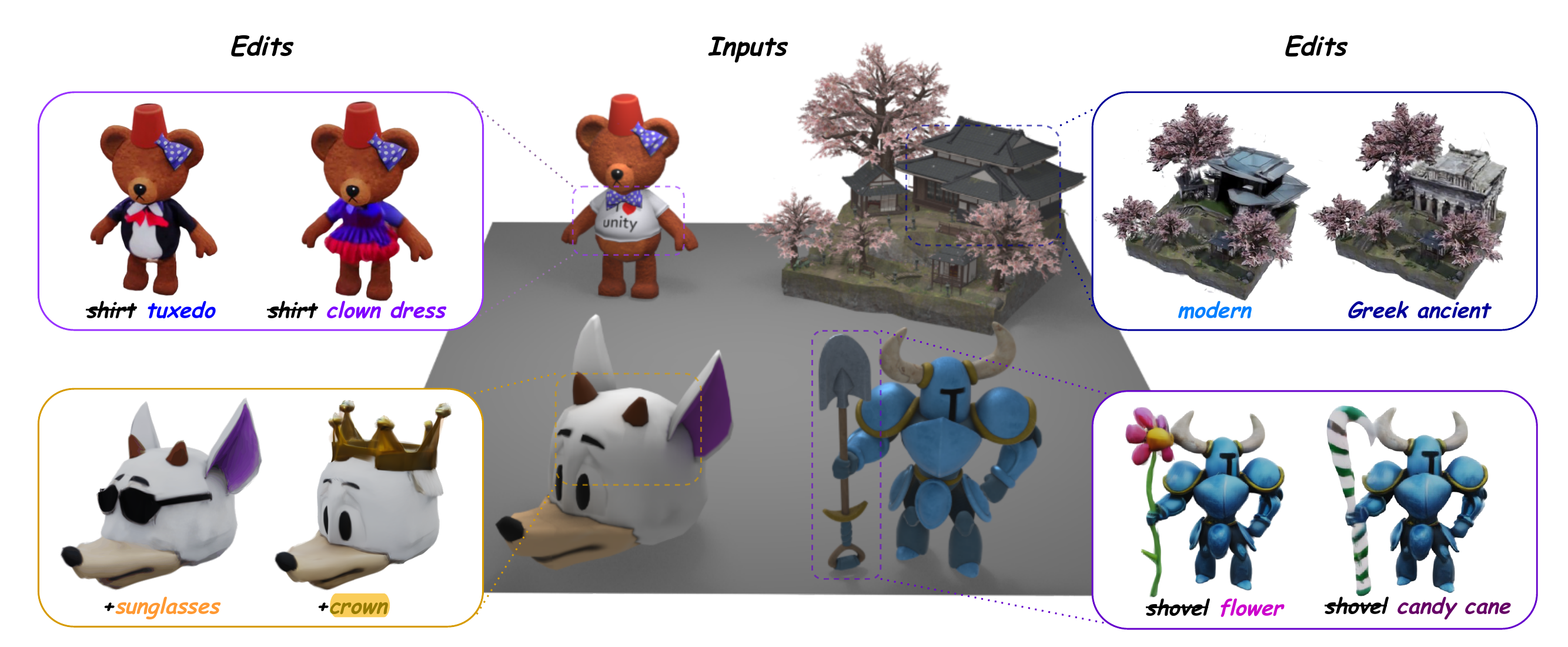}
    \captionof{figure}{
    \textbf{\name{}} takes a 3D  asset and a user-specified edit instruction as input and leverages the expressiveness of the latent space of a 3D diffusion model to generate edited 3D assets with high-quality details, geometric 3D consistency and precise adherence to diverse edit instructions.}
    \label{fig:teaser}
\end{center}
}]

\input{sec/0_abstract}    
\input{sec/1_intro}
\input{sec/2_related}
\input{sec/3_method}
\input{sec/4_experiments}
{
    \small
    \bibliographystyle{ieeenat_fullname}
    \bibliography{main}
}

\end{document}

%% file: sec/0_abstract.tex
\begin{abstract}
Despite the recent success of multi-view diffusion models for text/image-based 3D asset generation, instruction-based editing of 3D assets lacks surprisingly far behind the quality of generation models. The main reason is that recent approaches using 2D priors suffer from view-inconsistent editing signals. 
  Going beyond 2D prior distillation methods and multi-view editing strategies, we propose a training-free editing method that operates within the latent space of a native 3D diffusion model, allowing us to directly manipulate 3D geometry. We guide the edit synthesis by blending 3D attention maps from the generation with the source object. Coupled with geometry-aware regularization guidance, a spectral modulation strategy in the Fourier domain and a refinement step for 3D enhancement, our method outperforms previous 3D editing methods enabling high-fidelity and precise edits across a wide range of shapes and semantic manipulations.
  Our project webpage is: \href{https://mparelli.github.io/3d-latte}{https://mparelli.github.io/3d-latte}.

\end{abstract}

%% file: sec/1_intro.tex
\section{Introduction}
\label{sec:intro}

Advances in 2D diffusion models~\citep{rombach2021highresolution} and 3D representations, such as Neural Radiance Fields (NeRFs)~\cite{mildenhall2020nerf} and 3D Gaussian Splatting (3DGS)~\citep{kerbl3Dgaussians}, have revolutionized 3D asset creation. More recently, powerful text-to-3D~\cite{lin2025diffsplat,xiang2024structured} and image-to-3D~\citep{lin2025diffsplat, xiang2024structured, liu2023zero1to3, liu2023one2345++} generative models have emerged, enabling scalable and expressive 3D content creation by learning from large 3D data collections \cite{Deitke2023objaverse, objaverseXL}. A core challenge in this domain is instruction-based 3D editing: modifying the geometry and appearance of a 3D object based on a language instruction, while selectively preserving the object’s identity and structure. Achieving semantically accurate and high-fidelity 3D edits has attracted considerable attention due to its importance for applications in design and virtual and augmented reality. \par
Many methods propose to solve this task by distilling 2D diffusion priors~\citep{brooks2023instructpix2pix} into a 3D representation via score distillation losses~\citep{poole2022dreamfusion, li2023focaldreamer,sella2023vox,Koo2024PDS} or iterative dataset updates~\citep{Haque2023instructnerf, chen2023gaussianeditor, Chen2024proedit}. These approaches, however, are constrained by their reliance on 2D supervision. Specifically, they often exhibit multi-view inconsistencies, such as multi-face Janus issues, due to the limited 3D awareness, or inherit 2D editing failures from specific viewpoints, resulting in implausible 3D reconstructions. As a result, most existing methods succeed in appearance edits but struggle with large spatial or geometrical transformations that require globally consistent shape changes. A new line of work~\citep{erkoc2024preditor3d, li2025cmd,huang2025edit360} proposes to synchronously edit multiple views via multi-view diffusion priors or relies on feedforward 3D reconstruction models~\citep{zhuang2024gtr, LaRa, hong2023lrm} to consolidate them in 3D. Others adopt a hybrid 2D-3D approach~\citep{mvedit2024}, where multi-view images are fused into a 3D representation at each denoising step. Nevertheless, existing methods based on feedforward 3D reconstruction or multi-view priors often suffer from blurry, distorted reconstructions, due to small inconsistencies across views being propagated to the 3D model. Hybrid 2D-3D methods, on the other hand, may introduce Janus artifacts and multi-view inconsistencies due to the use of 2D priors at intermediate steps. \\
To overcome these limitations, we take an alternative approach and leverage a 3D-native diffusion prior, where noise is injected directly to the 3D representation. This allows us to directly manipulate the appearance and geometry of a 3D asset without reliance on 2D or multi-view priors. Importantly, we operate within the model's latent space, which is structured as pixel-aligned 3D Gaussian representations \citep{lin2025diffsplat}. Motivated by the role of attention maps in 2D editing~\citep{hertz2022prompt}, our key insight is that the 3D self- and cross-attention maps naturally capture information about the layout and composition of the 3D scene and directly model relationships between 3D Gaussians and text tokens. Building on this observation, we introduce a 3D attention injection mechanism into the denoising process of a text-conditioned 3D diffusion model. At each time step, we modulate the model’s attention maps by blending or replacing them with those obtained from the generation with the source prompt, namely the description of the original, unedited asset. This allows us to guide the model to synthesize a 3D asset that semantically aligns with the edit prompt while preserving the 3D structure of the source asset.\\ To achieve localized edits, we leverage a vision-language model (VLM) in combination with a segmentation model to generate multi-view consistent 2D masks. These masks naturally define a 3D segmentation over the multi-view pixel-aligned 3D Gaussians, allowing us to constrain the modification to the relevant 3D regions. To enhance 3D quality, we adopt a frequency-modulated strategy that emphasizes low-frequency components early in the denoising process, encouraging the model to capture global structure before refining fine-grained details. Structural coherence is further reinforced through a geometry-aware regularization term, applied in the form of classifier guidance. Finally, to address higher-fidelity edits we adopt an iterative strategy that progressively enhances the fine-grained details of the 3D representation while preserving cross-view consistency. \\
To demonstrate the effectiveness of our approach, we conduct extensive user studies, report quantitative metrics based on CLIP similarity~\citep{Haque2023instructnerf} and evaluate performance using GPTEval3D~\citep{wu2023gpteval3d}, consistently surpassing previous state-of-the-art works.


%% file: sec/2_related.tex
\vspace{-0.2cm}
\section{Related Works}
\label{sec:related}


\boldparagraph{3D Diffusion Generative Models} Recent methods in 3D generation have shifted their focus toward diffusion-based models~\citep{ho2020denoising}, which have been adapted to a plethora of 3D representations, including voxel grids~\citep{hui2022neural, trellis}, point clouds \citep{zeng2022lion,nichol2022pointe} and triplanes \citep{Shue2023TriDiff, anciukevicius2022renderdiffusion}.
Recently, methods leveraging 3DGS \citep{yang2024prometheus, Zhou2025diffgs, lin2025diffsplat,zhang2024gaussiancube} have emerged as particularly promising. GaussianCube~\citep{zhang2024gaussiancube} organizes a 3DGS representation into a voxel field and applies 3D diffusion using a 3D U-Net. In contrast, DiffSplat~\citep{lin2025diffsplat} introduces a 3D latent diffusion model that directly generates 3D Gaussians by modeling an object as a set of multi-view 3DGS grids.


\boldparagraph{3D Editing with 2D Priors}
Editing in 3D presents unique challenges due to the need for multi-view consistency. One line of work tackles this by leveraging image-conditioned 2D diffusion models such as InstructPix2Pix~\citep{brooks2023instructpix2pix}, injecting their editing capabilities into learned 3D representations. A pioneering work in this direction, InstructNeRF2NeRF~\citep{Haque2023instructnerf}, introduces the ``Iterative Dataset Update'' method, where rendered views from the 3D model are repeatedly updated during optimization. More recent methods~\citep{chen2023gaussianeditor, Chen2024proedit}, build upon this paradigm. However, large edits can still lead to multi-view inconsistencies due to the lack of explicit multi-view awareness. Another line of work, ~\cite{zhuang2023dreameditor, li2023focaldreamer, sella2023vox}, relies on Score Distillation Sampling (SDS) to guide 3D asset synthesis by distilling gradients from a pre-trained diffusion model. Posterior Distillation Sampling~\citep{Koo2024PDS} improves upon SDS by aligning the latents of source and target images. While effective, these methods inherit known limitations of SDS, including Janus artifacts and mode-seeking behavior.
As an alternative, recent methods~\citep{wu2024gaussctrl, lee2025editsplat, chen2024dge, wen2025intergsedit} perform multi-view consistent edits directly on source images, which are then used to update the 3D representation. For instance, DGE~\citep{chen2024dge} leverages epipolar constraints to aggregate features across views, while GaussCTRL~\citep{wu2024gaussctrl} performs depth-conditioned 2D updates combined with cross-view alignment. However, such reliance on depth guidance can prevent the model from performing significant shape changes. In contrast, by operating directly within the latent space of a 3D diffusion model, our method enables flexible manipulation of both appearance and geometry in a fully 3D-consistent manner without relying on 2D editing priors or SDS-based optimization. 

\boldparagraph{Hybrid 2D-3D Editing Methods}
In parallel, another line of work has emerged that moves beyond pure 2D updates by incorporating hybrid 2D-3D strategies. SHAP-Editor~\citep{chen2024shap} learns a feed-forward editor operating in the latent space of Shap-E~\citep{Jun2023ShapEGC} and is trained using an SDS~\citep{poole2022dreamfusion} objective. However, it requires retraining for each new set of edits and relies heavily on the 2D priors and latent structure of Shap-E~\citep{Jun2023ShapEGC}. Thus, its edits lack flexibility and are often of poor visual quality. In contrast, our method generalizes across categories, produces high-fidelity outputs, and enables fast inference. MVEdit~\citep{mvedit2024} adopts a hybrid 2D-3D approach by fusing images into a 3D representation between denoising steps in a multi-view diffusion model.  More recently, Edit360~\cite{huang2025edit360} builds on a dense-view synthesis model and achieves propagation of edits by aligning an
edited view camera trajectory with the original front-view trajectory. While these are promising steps toward enabling 3D-aware edits, reliance on 2D priors or trajectory alignment strategies can introduce multi-view inconsistencies.

%% file: sec/3_method.tex
\vspace{-0.2cm}
\section{Method}

\begin{figure*}[ht] 
    \centering
    \includegraphics[width=0.95\textwidth]{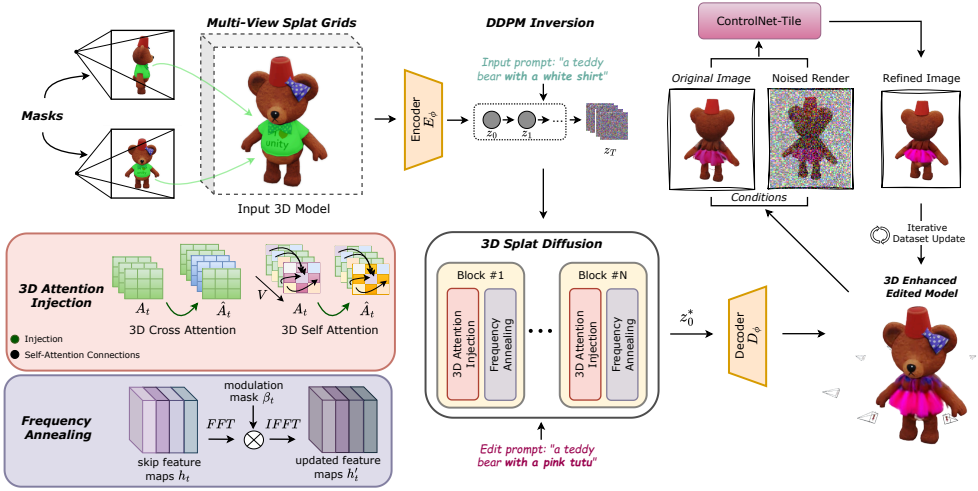} 
\caption{\textbf{Overview of \name{}}. We operate in the latent space of a pre-trained 3D generative model. The source 3D object is represented as a multi-view Gaussian splat grid and inverted into its corresponding noise latent. Starting from this latent, we perform denoising guided by the edit prompt, while injecting 3D cross- and self-attention maps derived from the source object. A geometry regularization guidance term, a frequency modulation strategy and a 3D enhancement module further refine the result. Region-specific edits are supported via masks generated using GroundingDINO~\citep{liu2023grounding} and SAM2~\citep{ravi2024sam2}.}
    \label{fig:method_figure}
    \vspace{-0.5cm}
\end{figure*}

Given a 3D object, a source text prompt \(p\) describing its original appearance and a target text prompt \( p^\ast \) describing the desired edit, our goal is to alter the object's appearance and/or geometry so that it semantically aligns with the target prompt. Simultaneously, we aim to preserve regions not referenced by the prompt and maintain the original object structure.
To this end, we introduce a zero-shot editing framework that extends attention control to the domain of 3DGS. Our method operates within the text-guided 3D diffusion model DiffSplat~\citep{lin2025diffsplat}, outlined in Section~\ref{sec:method.diffsplat}. Our core idea is to inject 3D cross- and self-attention maps derived from the source 3D asset during the diffusion process that synthesizes the edited 3D asset as explained in Section~\ref{sec:method.3dinject}. This enables semantically meaningful edits while maintaining multi-view consistency and 3D structural coherence, since our attention modifications are applied within a latent space that encodes 3D geometry. To further enhance quality, we incorporate a geometry-aware regularization mechanism (Section~\ref{sec:method.georegul}) and a frequency annealing strategy (Section~\ref{sec:method.freqanneal}). Finally, in Section~\ref{sec:method.3Denhancement} we introduce an iterative refinement pipeline that progressively enhances high-frequency details and texture fidelity in the reconstructed 3D asset. An overview of our method is shown in Figure~\ref{fig:method_figure}.\par

\subsection{3D Representation and 3D Diffusion Model}
\label{sec:method.diffsplat}

In this work, we leverage DiffSplat~\citep{lin2025diffsplat} as our 3D diffusion backbone. In~\citep{lin2025diffsplat} a 3D object is modeled as a set of structured multi-view splat grids \( \mathcal{G} = \{G_i\}_{i=1}^{V} \), where each \( G_i \in \mathbb{R}^{C \times H \times W} \), \( V \) is the number of input views, \( H \times W \) is the spatial resolution, and \( C = 12 \) corresponds to the number of Gaussian attributes. Each Gaussian primitive \( g_i \in \mathbb{R}^{12} \) is parameterized by its RGB color \( c_i \in \mathbb{R}^3 \), 3D location \( x_i \in \mathbb{R}^3 \), which is determined by its depth and camera parameters, scale \( s_i \in \mathbb{R}^3 \), rotation quaternion \( r_i \in \mathbb{R}^4 \), and opacity \( o_i \in \mathbb{R} \). The pipeline of~\citep{lin2025diffsplat} comprises a Gaussian reconstruction module that converts multi-view RGB images, along with depth and normal maps into a Gaussian splat grid representation, a VAE that encodes the Gaussian splat grid into a latent space and a generative model that is trained on this representation. Finally, the reconstructed latent is fed to a VAE decoder to obtain the 3D gaussian representation, which can be rendered as multi-view images.
\vspace{-0.1cm}
\subsection{3D Inversion}
\label{sec:method.inversion}
Following \citep{lin2025diffsplat}, we represent a 3D object as a set of structured multi-view splat grids \( \mathcal{G} = \{G_i\}_{i=1}^{V} \).
We use the reconstruction module from DiffSplat~\citep{lin2025diffsplat}, which takes as input a set of \( V \) RGB images, depth maps, and normal maps rendered from uniformly distributed camera viewpoints around the source 3D object, resulting in a 3D Gaussian for each pixel of the input views. 
We begin by inverting the source 3D object representation into its corresponding noise latent \( z_T \). However, standard DDIM inversion~\citep{song2021ddim} can introduce slight errors, leading to poor reconstructions. Thus, we adapt the DDPM inversion mechanism proposed in ~\citep{huberman2024edit} to operate on a set of multi-view Gaussian splat grids \( \mathcal{G} \) and recover a noise vector that accurately reconstructs the original 3D object.\\
Let the encoded Gaussian splat grids (i.e., splat latents) be denoted as $z_0 = \{ z_{0}^{i} \}_{i=1}^{V} = \{E_{\phi}(G_i)\}_{i=1}^{V}$
 where ${E_{\phi}}(\cdot)$ is the Gaussian VAE encoder of \citep{lin2025diffsplat}. 
We construct the forward diffusion trajectory~\citep{song2021ddim} and obtain the noise  maps $\eta_t$ as follows:
\vspace{-0.2cm}
\begin{equation}
z_t = \sqrt{\alpha_t}\, z_0 + \sqrt{1 - \alpha_t}\, \epsilon_t,
\quad
\eta_t = \frac{(z_{t-1} - \mu_\theta(z_t, t))}{\sigma_t}
\end{equation}
where $\alpha_t$ is the cumulative diffusion noise schedule, 
$\epsilon_t \sim \mathcal{N}(0,I)$ is sampled independently per timestep, 
and $\mu_\theta$ and $\sigma_t$ are the predicted mean and standard deviation 
of the noise injected at each timestep, respectively.
This sequence of noise maps is shown in \citep{huberman2024edit} to be edit-friendly since they are constructed with statistically independent noise samples.
Starting from \( z_T \), we perform the denoising process with the edit prompt $p^\ast$. To achieve editing while maintaining 3D structural information and layout, we apply 3D self- and cross-attention injection. 
\vspace{-0.1cm}
\subsection{3D Attention Injection}
\label{sec:method.3dinject}
\begin{figure}
\centering
\includegraphics[width=0.4\textwidth]{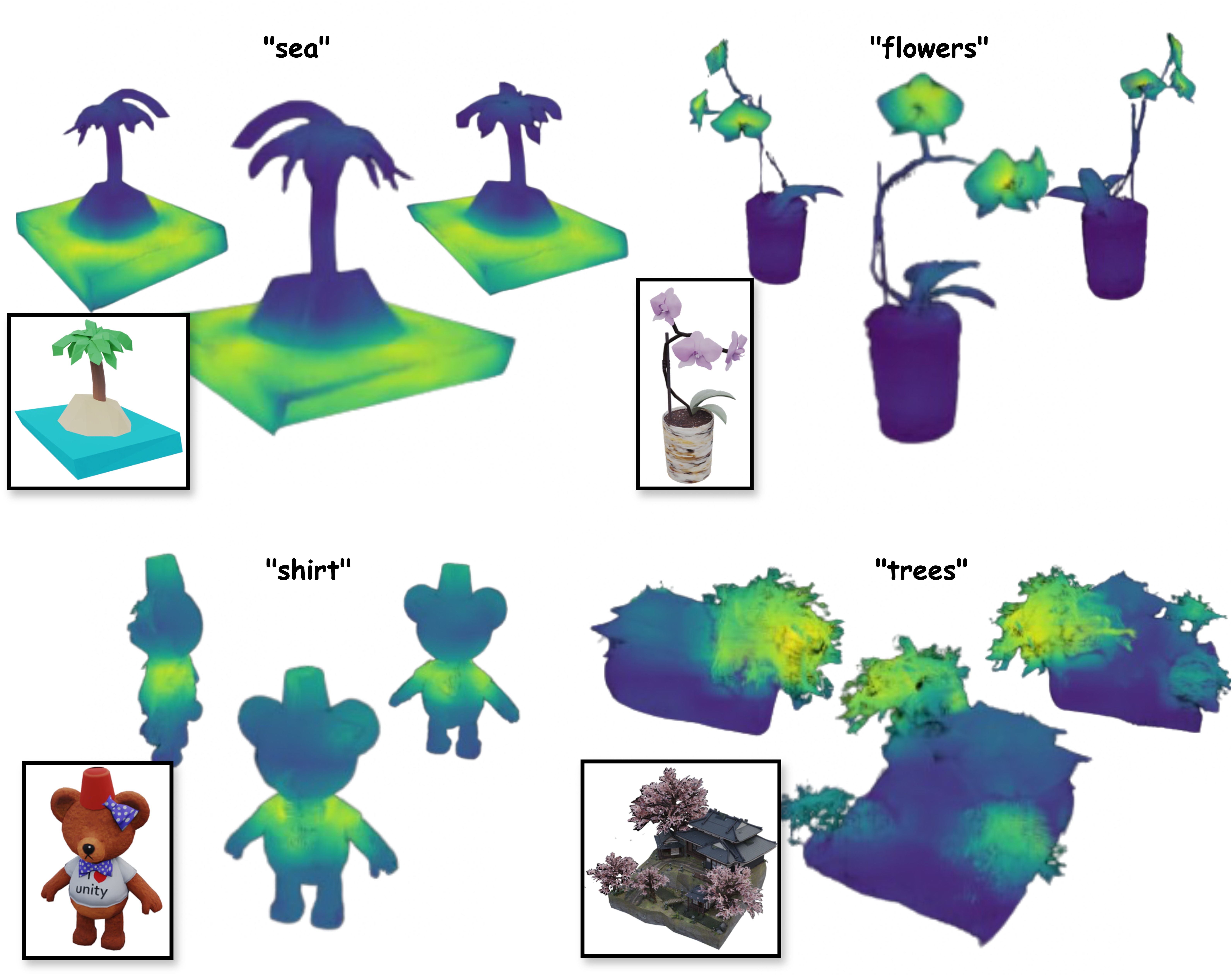} 
    \caption{Per-token attention over 3D splats, rendered from multiple viewpoints.}
    \label{fig:attention_viz}
    \vspace{-0.5cm}
\end{figure}

In our formulation, 3D-aware attention refers to the \emph{self}- and \emph{cross-attention maps} over the noisy multi-view Gaussian latents. In our 3D diffusion backbone at each timestep, features of the noisy Gaussian splat latents $\phi({z}_t) \in \mathbb{R}^{V \times D \times H \times W}$ are projected into a query matrix, $Q$, while the keys, $K$, and values, $V$, are obtained from the text prompt embeddings. Each entry $W_{i,j}$ in the 3D cross-attention map $ W_{\mathcal{G}}^{\text{cross}}$ represents the influence of the $j$-th text token on the $i$-th Gaussian latent. As illustrated in Figure~\ref{fig:attention_viz}, cross-attention weights define a token-3D Gaussian splat correspondence field, enabling accurate 3D localization of interest regions. Self-attention maps $W_{\mathcal{G}}^\text{self}$ capture spatial and semantic relationships between all 3D Gaussian latents.
Queries, keys and values are computed via linear projections of the Gaussian features.\\
Motivated by the role of attention in encoding 3D structure and semantics within the 3D Gaussian splat representation, we introduce a 3D attention injection framework.  Let \( z_{t-1} = D_{\theta}(z_t, t, p) \) denote the denoising step with the source prompt $p$, and \( z^\ast_{t-1} = D_{\theta}(z_t^*, t, p^*) \) be the denoising step with the edit prompt \( p^*\).
Additionally, let ${W_{\mathcal{G}}}_{t} = ({W_\mathcal{G}}^{\text{cross}}_t, {W_\mathcal{G}}_t^{\text{self}})$ and ${W_\mathcal{G}}_t^\ast = (({W_\mathcal{G}}_t^\ast)^{\text{cross}}, ({W_\mathcal{G}}_t^\ast)^{\text{self}})$ be the attention maps computed in the denoising steps \( D_{\theta}(z_t, t, p) \) and \( D_{\theta}(z_t^*, t, p^*) \), respectively.
We inject modified attention maps \( \hat{W_\mathcal{G}}_t = (\hat{W_\mathcal{G}}_t^{\text{cross}}, \hat{W_\mathcal{G}}_t^{\text{self}}) \), derived from the source prompt \( p \), into the denoising process guided by the edit prompt \( p^\ast \), as follows.
At each denoising step with prompt \( p^\ast \), we override the attention maps ${W_{\mathcal{G}}}^\ast_t$ with $\hat{W_\mathcal{G}}_t$ (i.e., ${W_\mathcal{G}}^\ast_t \leftarrow \hat{W_\mathcal{G}}_t$).
To calculate the modified cross-attention map, we follow \citep{hertz2022prompt} and inject 3D attention until a timestep $\tau_{\text{cross}}$ only on tokens shared between the original and edited prompts, using the following $F_{cross}^{3D}$ function:
\begin{equation}
\begin{aligned}
\hat{W_\mathcal{G}}_t^{\text{cross}} &=
F_{\text{cross}}^{3D}\!\left(
W_{\mathcal{G}}^{\text{cross}}, (W_{\mathcal{G}}^\ast)^{\text{cross}}, t
\right)_{i,j} \\[-2pt]
&=
\begin{cases}
(({W_\mathcal{G}}_t^\ast)^{\text{cross}})_{i,j}, &
\text{if } CT(j)=\varnothing \text{ or } t<\tau_{\text{cross}},\\[4pt]
({W_\mathcal{G}}_t^{\text{cross}})_{i,CT(j)}, &
\text{otherwise.}
\end{cases}
\end{aligned}
\label{eq:cross-weight-update}
\end{equation}
where $CT$ is an alignment function that maps a token index from $p^\ast$ to the corresponding token index in $p$. In the case where there is no match, it returns an empty set.
 \begin{figure}[ht]
 \centering
\includegraphics[width=0.39\textwidth, trim={30. 10 30 10},clip]{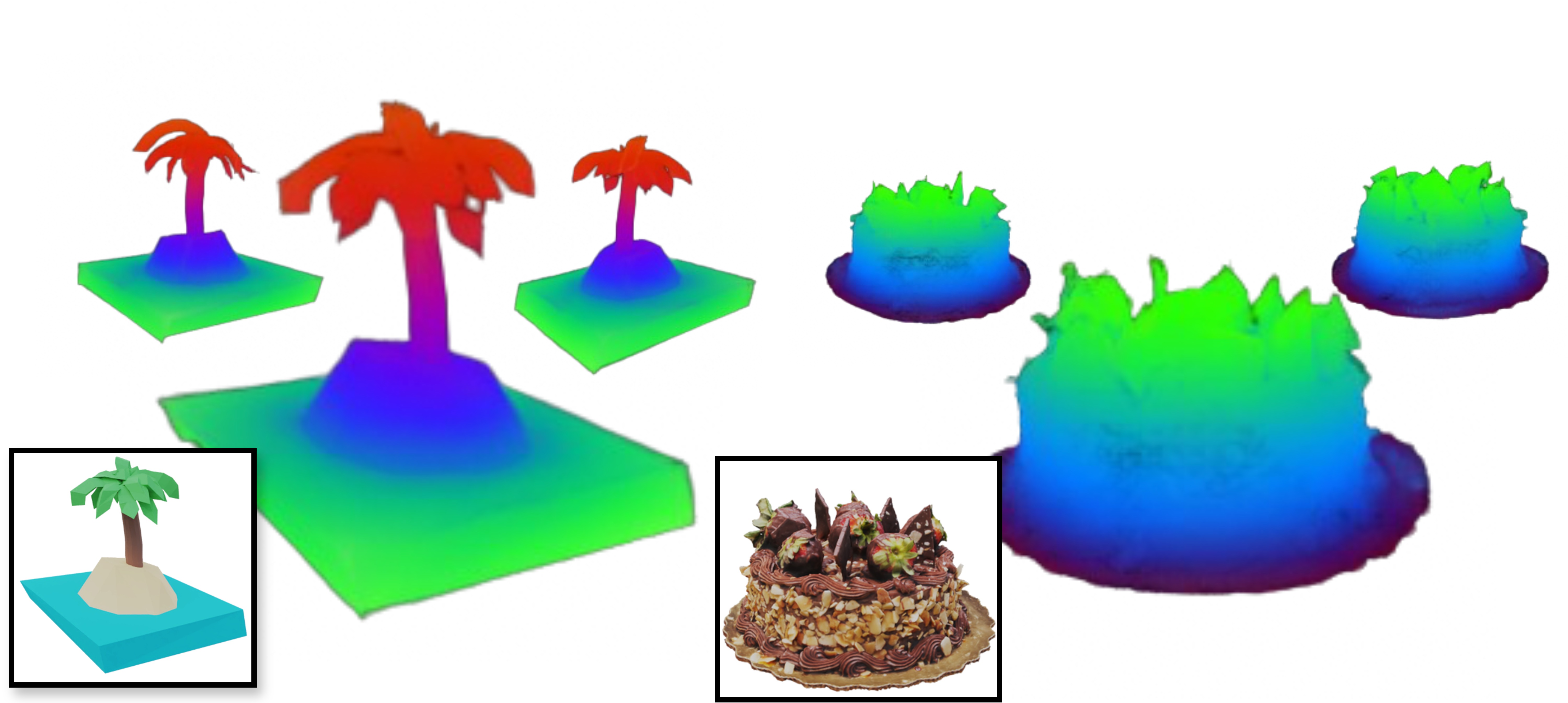} 
    \caption{We form the normalized Laplacian of the 3D self-attention graph and map the first three eigenvectors to a color value per Gaussian.}
    \label{fig:self_attention_viz}
    \vspace{-0.4cm}
\end{figure}
By also injecting 3D self-attention during the early timesteps of the denoising process, we encourage the model to further preserve 3D structural relationships of the source asset, such as part layout, composition and spatial symmetry, before the introduction of semantic details. As illustrated in Figure~\ref{fig:self_attention_viz}, the spectral decomposition of the 3D self-attention graph reflects 3D scene composition and groups 3D Gaussians into distinct semantic parts.
In the case of self-attention, the $F_{self}^{3D}$ function is defined as follows:
\begin{equation}
\begin{aligned}
\hat{W_\mathcal{G}}_t^{\text{self}} &=
F_{\text{self}}^{3D}\!\left(
W_{\mathcal{G}}^{\text{self}}, (W_{\mathcal{G}}^\ast)^{\text{self}}, t
\right) \\
&=
\begin{cases}
({W_\mathcal{G}}_t^\ast)^{\text{self}}, & 
\text{if } t < \tau_{\text{self}},\\
{W_\mathcal{G}}_t^{\text{self}}, & 
\text{otherwise.}
\end{cases}
\end{aligned}
\label{eq:self-weight-update}
\end{equation}
where $\tau_{\text{self}}$ is the timestep until which injection is applied.
\vspace{-0.1cm}
\subsection {Mask Generation and Region Editing}

To extract the target edit area, we formulate a query to a large vision-language model (GPT-4o) that includes the original prompt, the edit instruction, and a rendered front-view of the object. The model is asked to identify which part of the object is affected by the edit (e.g., ``the shirt of the teddy bear''). We use these identified editing regions to generate 2D multi-view consistent segmentation masks. More specifically, GroundingDINO~\citep{liu2023grounding} processes the multi-view rendered images and the edit regions, proposed by GPT-4o to generate bounding boxes, which are subsequently refined into 2D segmentation masks tracked by SAM 2~\citep{ravi2024sam2}. This process is outlined in the supplementary.
Benefiting from our pixel-aligned representation, these masks naturally approximate a 3D segmentation of the corresponding pixel-aligned Gaussians. To allow for flexible geometric modifications, we approximate the initial 3D mask expansion by computing the cross-attention map \( \overline{{W_\mathcal{G}}_{t}^{r^\ast}} \) averaged over all time steps for the prompt token \( r^\ast \) that refers to the desired editing region (e.g., ``tutu'' in Figure~\ref{fig:method_figure}), and applying a threshold to obtain binary masks that indicate 3D regions influenced by the new concept. The final 3D editing region $M$ is defined as the union of the original masks predicted by SAM 2 lifted to 3D and the attention-derived mask. We constrain the diffusion process to this area, by blending the original latent $z_{t-1}$ with the edited latent $z_{t-1}^*$ as follows:
\begin{equation}
\hat{z}_{t-1} = (1 - M) \odot z_{t-1} + M \odot z_{t-1}^*
\label{eq:blend}
\end{equation}
where $\odot$ denotes element-wise multiplication. Our method also supports user-defined masks.
\vspace{-0.1cm}
\subsection{Geometry Regularization}
\label{sec:method.georegul}
The editing signal via attention injection can introduce some uncertainty in the editing regions, resulting in Gaussians prone to semi-transparencies and premature collapse. We find that a simple geometric regularizer can help circumvent these artifacts. Thus, we incorporate a soft geometry-aware classifier guidance mechanism that softly penalizes the removal of Gaussians in regions that are highly relevant to the edit. To this end, we compute a soft mask $R_t^i \in [0,1]$ for each Gaussian $i$, which reflects how relevant that Gaussian is to the current edit at denoising step $t$. We define the editing signal at each timestep as the L1 difference between the noise predictions conditioned on the edited and original prompts:
$
D^i = \left\| \epsilon_\theta(z_t, t, p^\ast) - \epsilon_\theta(z_t, t, p) \right\|_1 ,$
where $\epsilon_\theta(z_t, t, p)$ and $\epsilon_\theta(z_t, t, p^\ast)$ denote the denoising model's noise predictions under the original and edited prompts, respectively. To obtain a soft relevance mask $R_t$ we normalize the absolute difference values $D^i$ across all Gaussians via global min--max normalization.
Thus, $R_t^i$ denotes the relevance weight of the $i$-th Gaussian, arranged in a pixel-aligned grid of size $V \times H \times W$. Our intuition is that the magnitude of this mismatch reflects the relevance of each 3D Gaussian to the edit. As the existence of Gaussians is determined by their opacity and covariance, our regularization loss is defined as:
\begin{equation}
\resizebox{0.95\hsize}{!}{$
\mathcal{L}_{\text{geo}} =
\lambda_o \sum_i R_{t}^{i} \cdot \exp(-\gamma_o \cdot o_i)
+
\lambda_\Sigma \sum_i R_{t}^{i} \cdot \exp(-\gamma_\Sigma \cdot \operatorname{Tr}(\Sigma_i))
$}
\end{equation}
where $o$ and $\Sigma$ are obtained by decoding the denoised gaussian splat latents with the VAE decoder $D_{\phi}$ from~\citep{lin2025diffsplat}. Here, $\gamma_o$ and $\gamma_\Sigma$ control the sharpness of the penalties, and $\lambda_o$, $\lambda_\Sigma$ weigh the opacity and scale terms. This formulation softly penalizes low opacity and insufficient spatial support. We incorporate our geometry-aware regularization loss as a guidance signal during denoising:
\begin{equation}
z_{t-1} = \hat{z}_{t-1} - s \cdot \nabla_{{z}_t} \mathcal{L}_{\text{geo}}({z}_t)
\end{equation}
where $s$ denotes the guidance scale, and $\hat{z}_{t-1}$ is the predicted latent from Equation~\ref{eq:blend}. 
\vspace{-0.1cm}
\subsection {Frequency Annealing}
\vspace{-0.1cm}
\label{sec:method.freqanneal}
The injection of attention from the source prompt during editing can impact the model's denoising ability, intensifying the misalignment between the condition and diffusion spaces. In some examples, this causes the model to overemphasize high-frequency visual features of the source, such as decorative textures. These may become partially preserved and degrade into artifacts on the surface of the generated 3D asset. 
Consistent with observations in 2D~\citep{wu2024freeinit, si2023freeu}, in 3D generation, high-frequency components correspond to fine details while low-frequencies primarily capture 3D geometric structure and semantic layout.
Thus, we introduce a frequency annealing strategy, where we apply spectral modulation in the Fourier domain of the feature maps from the U-Net skip connections at each denoising step. 
Given the gaussian feature map $h_{l,t} \in \mathbb{R}^{V \times C \times H \times W}$ at layer $l$ and timestep $t$, we apply the following steps:
\begin{equation}
\resizebox{\hsize}{!}{$
\begin{alignedat}{2}
&\begin{aligned}
F(h_{l,t}) &= \mathrm{FFT}(h_{l,t}) \\
F'(h_{l,t}) &= F(h_{l,t}) \odot \beta_{l,t} \\
h'_{l,t} &= \mathrm{IFFT}(F'(h_{l,t}))
\end{aligned}
&& \quad
\begin{aligned}
\beta_{l,t}(r) &=
\begin{cases}
s_l, & \text{if } t > \tau \text{ and } r < r_{\text{thresh}},\\[2pt]
s_h, & \text{if } t \leq \tau \text{ and } r \geq r_{\text{thresh}},\\[2pt]
1,   & \text{otherwise.}
\end{cases}
\end{aligned}
\end{alignedat}
\label{eq:fft_beta}
$}
\end{equation}

Here, $\text{FFT}(\cdot)$ and $\text{IFFT}(\cdot)$ denote the Fourier transform and its inverse. The modulation mask $\beta_{l,t}$ applies scaling based on the radius $r$ and the terms $s_l$ and $s_{h}$ denote the scaling factors applied to the low- and high-frequency components respectively. This encourages the model to boost low-frequency components during early denoising with attention injection and then transition to emphasize high-frequency components, enabling structural preservation of the source without unwanted oversmoothed textures.
\vspace{-0.1cm}
\subsection{3D Enhancement}
\vspace{-0.1cm}
\label{sec:method.3Denhancement}
Many existing text-to-3D generation models, particularly those fine-tuned on datasets with plain surface details, are constrained to lower-resolution outputs and struggle to capture fine-grained geometric and appearance details. Thus, in challenging cases involving finer structures when rendering our 3D Gaussian representation at higher resolutions, we observe a degradation in appearance quality, characterized by blurred textures and limited details.
Drawing inspiration from~\citep{Haque2023instructnerf} to improve visual fidelity while preserving 3D structure and consistency, we leverage an iterative dataset update technique. Unlike prior work that primarily uses such pipelines for editing~\citep{Haque2023instructnerf, chen2023gaussianeditor}, we adapt them for 3D enhancement. Specifically, we follow an iterative process that comprises three key steps: i) rendering high-resolution views from the edited 3DGS representation, ii) enhancing these views by inputting noised version of them to the 2D diffusion backbone model, conditioned on the original edited images and, (iii) re-optimizing the 3DGS model with the updated enhanced images. To constrain the extent of correction to the editing area, the updated image is represented as:
$
I_{\text{blend}} = M \odot I_e + (1 - M) \odot I_{\text{src}},
$
where \( M \) denotes the mask containing the changed region, \( I_e \) denotes the enhanced image and \( I_{\text{src}}\) denotes the image of the source object. This process is shown in the supplementary.\\
As our 2D enhancement backbone, we leverage the ControlNet-Tile model~\cite{zhang2023adding}, which is designed for structure-preserving super-resolution, and is effective at restoring high-frequency details. By progressively re-rendering and updating the images we converge to a globally consistent higher-fidelity 3D representation.

%% file: sec/4_experiments.tex
\vspace{-0.1cm}
\section{Experiments}
\vspace{-0.1cm}
\label{sec:experiments}

\begin{figure*}[ht]
    \centering
    \includegraphics[width=\textwidth, trim = 40 320 40 80, clip]{gfx/qual_comp_edit.jpg} 
    \caption{\textbf{Qualitative comparison with baselines.} Our approach achieves the most plausible edits wrt. the input instruction text, while preserving the unedited parts of the 3D objects.}
    \label{fig:qual_results}
    \vspace{-0.5cm}
\end{figure*}
\begin{table}[ht]
\centering
\resizebox{\columnwidth}{!}{
\begin{tabular}{lccc}
\toprule
\textbf{Method} & \textbf{CLIP Dir} ↑  & \textbf{CLIP Diff No-Edit} ↓ & \textbf{CLIP-Dir-Con} ↑ \\
\midrule
MVEdit~\cite{mvedit2024}     & 0.121  & 0.077 & \third{0.67}\\
Vox-E~\cite{sella2023vox}      & \third{0.129}  & 0.054 & \second{0.68}\\
GaussCTRL~\cite{wu2024gaussctrl}  & 0.076  & \best{0.035} & 0.61\\
Edit360~\cite{huang2025edit360} & \second{0.149} & \third{0.045} & 0.59\\
PDS~\citep{Koo2024PDS} & 0.051& 0.094 & 0.55 \\
InstructGS2GS~\cite{igs2gs} &0.069 & 0.082 & 0.64\\
3D-LATTE (Ours)       & \best{0.178} & \second{0.039} & \best{0.77} \\
\bottomrule
\end{tabular}
}
 \caption{\textbf{Quantitative comparison using CLIP score metrics}. \bestcap{Best}, \secondcap{second-best} and \thirdcap{third-best} results are indicated.}
\label{tab:clip_directional_metrics}
\vspace{-0.4cm}
\end{table}

To evaluate our method, we construct a benchmark of 25 diverse 3D assets, each paired with a set of distinct edit instructions, resulting in a total of 100 samples. The assets are sourced from the Objaverse~\citep{Deitke2023objaverse} and Google Scanned Objects (GSO)~\citep{Downs2022GSO} datasets.
As main baselines, we include Vox-E~\citep{sella2023vox}, a voxel-based method using SDS, MVEdit~\citep{mvedit2024}, GaussCTRL~\cite{wu2024gaussctrl} and Edit360~\citep{huang2025edit360}. 
For completeness, we also report quantitative results on two additional  well-studied baselines: InstructGS2GS~\citep{igs2gs} and Posterior Distillation Sampling (PDS)~\citep{Koo2024PDS}. For the quantitative evaluation, we adopt the CLIP directional similarity (CLIP-dir) metric~\citep{Haque2023instructnerf} to quantify semantic alignment to edits by measuring direction changes in CLIP text embeddings and corresponding image embeddings of multi-view renderings.
To measure how well unedited regions are preserved, we include the CLIP-diff-no-edit metric~\citep{erkoc2024preditor3d},  which measures the CLIP score difference between input and output images using a modified text prompt where the edited part is replaced with a generic placeholder. Lower values indicate better shape preservation. Finally, to quantitatively evaluate if the edit propagates consistently across viewpoints we adopt the CLIP direction consistency score (CLIP-Dir-Con)~\cite{Haque2023instructnerf}, which computes how much the editing directions differ across frames. More details about the metrics are presented in the supplementary. To complement these metrics, we report results from GPTEval3D~\citep{wu2023gpteval3d}, a GPT-4V-based evaluation protocol in which the model compares multi-view renderings along three criteria: Text Prompt–3D Alignment, 3D Plausibility, and Texture Details.

\vspace{-0.1cm}
\subsection{Qualitative Results}
\vspace{-0.1cm}
As illustrated in Figure~\ref{fig:qual_results_extra} and \ref{fig:qual_results}, our method produces high-quality, multi-view consistent edits that are faithful to the input prompt. It is capable of handling significant geometric transformations, such as turning a shovel into a flower, or a house into an ancient building while preserving unedited regions and the structural integrity of the original object. As shown in Figure~\ref{fig:qual_results}, MVEdit~\citep{mvedit2024} successfully modifies appearance but struggles with morphological changes and suffers from multi-view inconsistencies. GaussCTRL~\citep{wu2024gaussctrl} exhibits similar limitations as it also struggles with geometric transformations, due to its reliance on depth-guidance. 
Vox-E~\citep{sella2023vox} handles geometric changes more robustly but suffers from low quality and sometimes fails to localize edits (e.g., wings on the character). Moreover, it is prone to unnatural 3D geometries due to the limited 3D awareness of SDS. Edit360~\cite{huang2025edit360} shows promise in edit faithfulness, however, the fusion of front and edited view trajectories introduces significant multi-view inconsistencies. Additional comparisons can be found in the supplementary.
\begin{figure*}
    \centering
    \includegraphics[width=0.9\textwidth, trim = 40 120 40 80, clip]{gfx/qual_fig_smaller.jpg} 
    \caption{\textbf{Qualitative Results.} Our method yields high-quality 3D objects for a diverse set of edits.}
    \label{fig:qual_results_extra}
    \vspace{-0.4cm}
\end{figure*}

\vspace{-0.2cm}
\subsection{Quantitative Results}
\begin{figure}
\centering
\includegraphics[width=0.3\textwidth]
{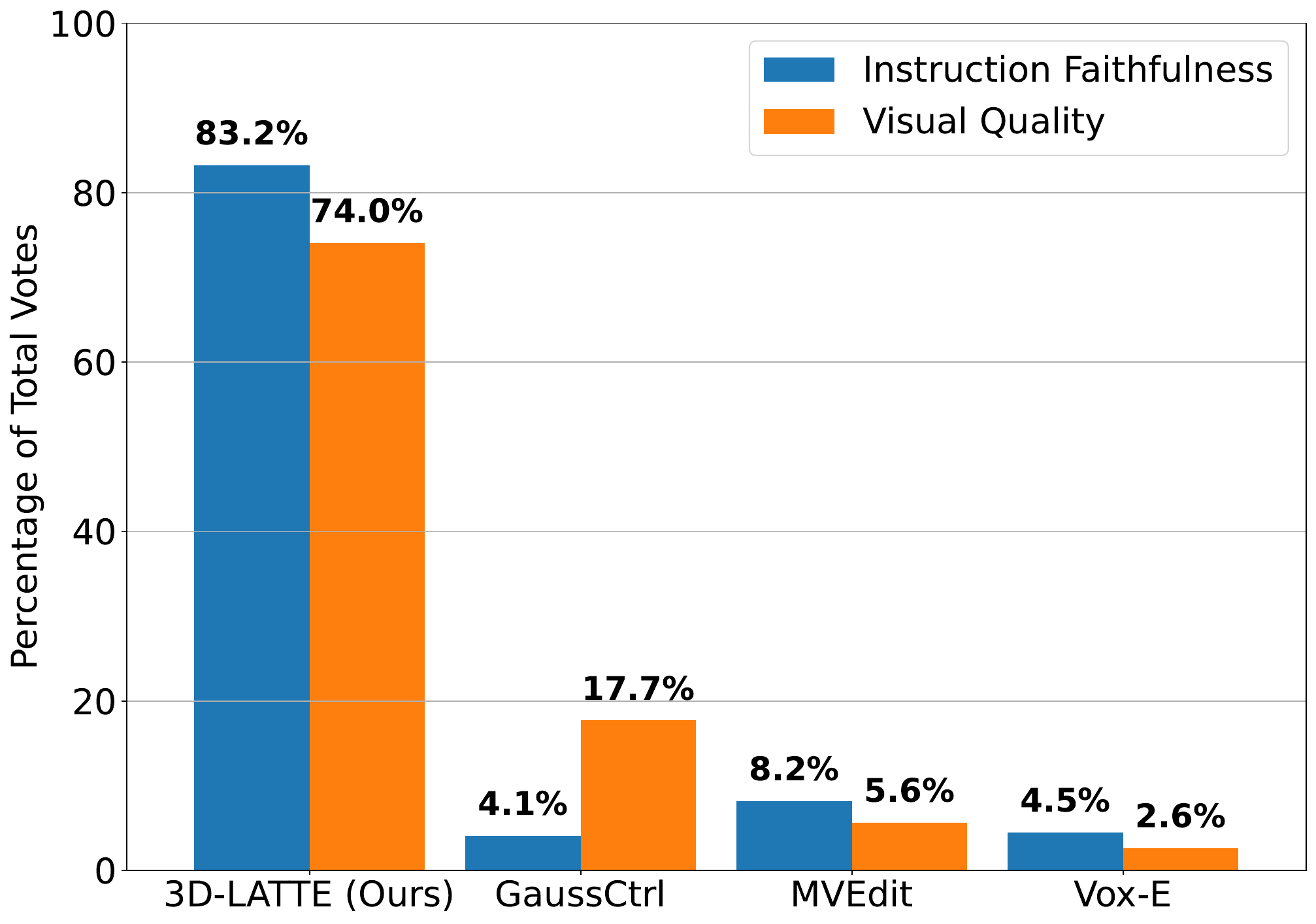} 
\vspace{-0.1cm}
    \caption{\textbf{User study.} Our approach shows a significantly higher percentage of votes in instruction faithfulness and visual quality.}
    \label{fig:user_study_results}
    \vspace{-0.2cm}
\end{figure}

\vspace{-0.1cm}
The quantitative comparison in Table~\ref{tab:clip_directional_metrics} supports the qualitative findings that our method surpasses the baselines
in terms of semantic alignment of predictions to the edit prompts measured by the CLIP-dir score.
The CLIP-no-diff score demonstrates that we preserve shapes effectively, compared to the other methods and we achieve a balanced tradeoff between editing and shape preservation. While GaussCTRL achieves a lower Diff-No-Edit score, this can be attributed to its frequent failure to apply the edits, leaving the objects unaltered. This is reflected in its significantly lower CLIP-dir score and in the qualitative examples.
In Table~\ref{tab:gpteval3d}, we show the results of the GPTEval3D~\citep{wu2023gpteval3d} metric in which GPT-4V shows a distinct preference for our method. \\
We also conduct a user study with 57 participants, who were asked to choose the best editing results among ours and three baselines, based on two criteria: visual quality and faithfulness to the edit prompt. The results in Figure~\ref{fig:user_study_results} show that our method was preferred by a significant margin.

\begin{table}[t]
\centering
\vspace{-0.2cm}
\resizebox{\linewidth}{!}{
\begin{tabular}{lccc}
\toprule
\textbf{Method} & \textbf{Prompt Algn.} ↑ & \textbf{3D Plausibility} ↑ & \textbf{Texture} ↑ \\
\midrule
MVEdit~\cite{mvedit2024} & 87\% & 71\% & 70\% \\
Vox-E~\cite{sella2023vox} & 78\% & 81\% & 78\% \\
GaussCTRL~\cite{wu2024gaussctrl} & 94\% & 83\% & 81\% \\
Edit360~\cite{huang2025edit360} & 67\% & 90\% & 72\% \\
\bottomrule
\end{tabular}
}
\vspace{-0.2cm}
\caption{\textbf{Quantitative comparison using GPTEval3D.}
Following \cite{erkoc2024preditor3d}, the scores represent the percentage of comparisons where our method wins over the respective baselines.}
\label{tab:gpteval3d}
 \vspace{-0.6cm}

\end{table}
\vspace{-0.3cm}
\subsection{Ablation study}
\begin{figure}[ht]
\centering
\includegraphics[width=0.35\textwidth, trim = 20 20 20 20, clip]
{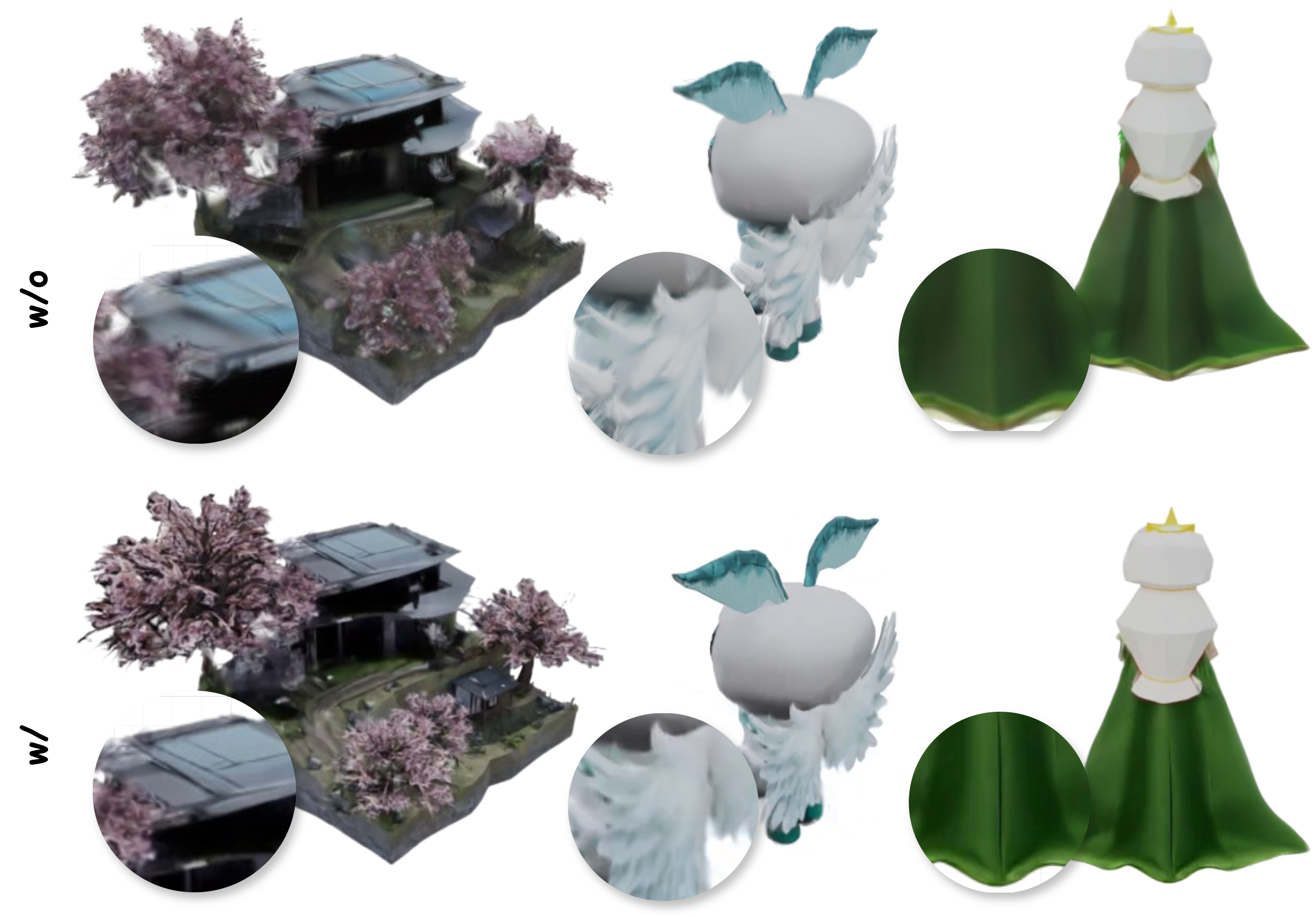} 
    \caption{\textbf{Effect of 3D enhancement.}}
    \label{fig:enhancement}
    \vspace{-0.4cm}
\end{figure}

We illustrate visual results showing the impact of the core components of our pipeline.  Additional quantitative results are included in the supplementary material. \\
\textbf{Effect of 3D Enhancement.} Figure~\ref{fig:enhancement} showcases the effect of our proposed 3D enhancement module. The top/bottom rows show a rendered view of the 3D objects before and after enhancement, demonstrating that the enhancement module recovers fine details and sharpens textures while maintaining 3D geometry(e.g., sharper architectural details).

\begin{figure}
    \centering
    \includegraphics[width=0.44\textwidth, trim = 20 20 20 20, clip]{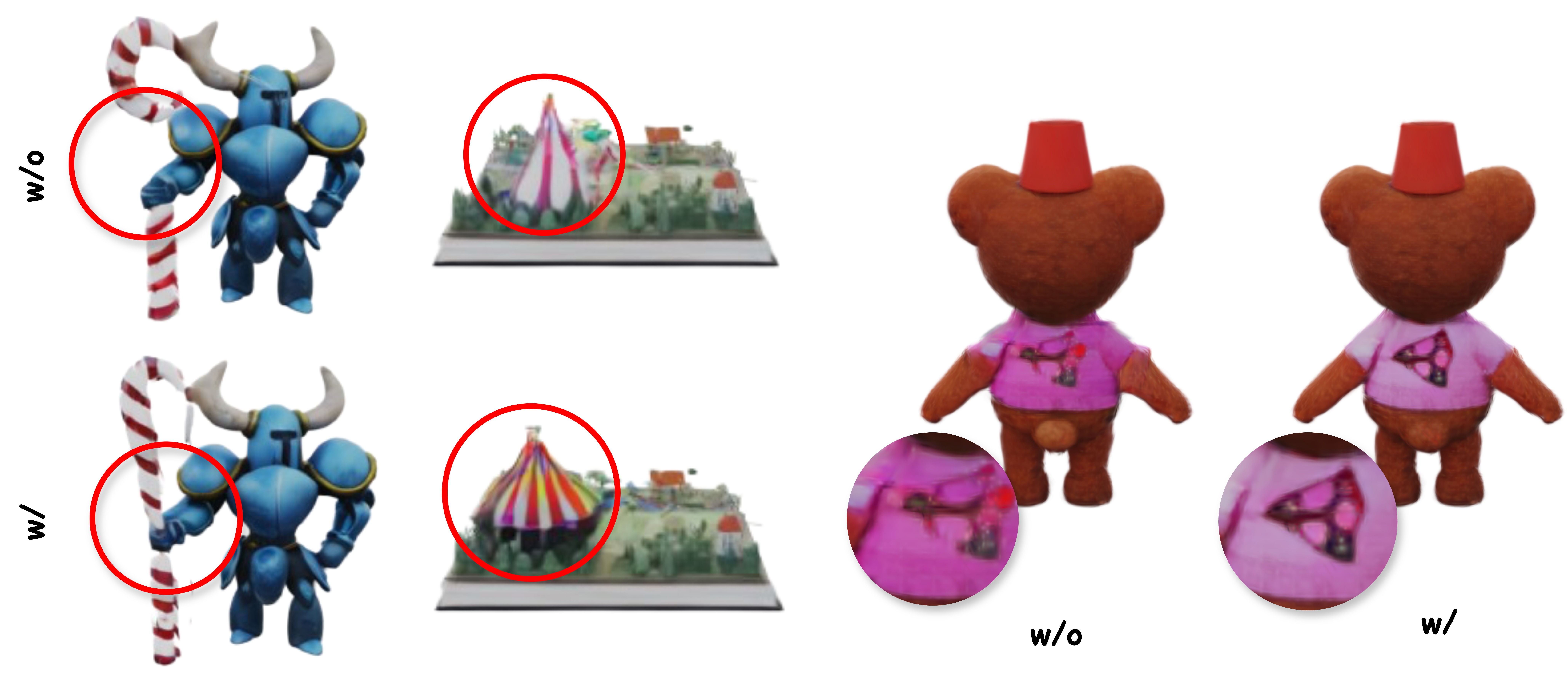}
    \parbox{0.49\textwidth}{
        \centering
         \begin{tabular}{@{\hspace{0.cm}}p{4.4cm}@{\hspace{0.2cm}}p{3.6cm}}
            (a) Geometry Regularization. & (b) Frequency Annealing.
        \end{tabular}
    }
            \vspace{-0.3cm}

    \caption{\textbf{Effect of our proposed additional components.}}
    \label{fig:ablation_combined}
        \vspace{-0.35cm}

\end{figure}

\textbf{Effect of Geometry Regularization}. We also qualitatively demonstrate the effectiveness of our geometry regularization guidance term. As shown in Figure~\ref{fig:ablation_combined}(a), without this term, some edited regions become partially transparent or vanish entirely, leading to degraded geometry. Our regularization mitigates these issues, resulting in robust edits. 

\textbf{Effect of Frequency Annealing.} As shown in Figure~\ref{fig:ablation_combined}(b), without frequency annealing, complex patterns in the source object, such as logos or prints that contain high-frequency information, can be overemphasized by the model, resulting in noisy textures in the final edit. 

\vspace{-0.3cm}
\section{Conclusion}
\vspace{-0.1cm}
In this paper, we introduced \name{}, a novel approach for instruction-based 3D editing. In contrast to existing approaches that rely on 2D distillation or multi-view priors, we leverage the expressiveness of the latent space of a native 3D diffusion model and perform editing by blending the respective 3D attention maps of the source and target object. Combined with geometry regularization, frequency annealing, and a 3D refinement step, our approach outperforms competing methods in both qualitative and  quantitative evaluations and user studies.

%% file: main.bib
@String(CVPR= {IEEE Conf. Comput. Vis. Pattern Recog.})

@String(ICCV= {Int. Conf. Comput. Vis.})

@String(ECCV= {Eur. Conf. Comput. Vis.})

@String(NIPS= {Adv. Neural Inform. Process. Syst.})

@String(TOG= {ACM Trans. Graph.})

@String(ICLR = {Int. Conf. Learn. Represent.})

@String(ICRA = {IEEE Int. Conf. Robot. Autom.})

@String(ICRA = {ICRA})

@String(CVPR  = {CVPR})

@String(ICCV  = {ICCV})

@String(ECCV  = {ECCV})

@String(NIPS  = {NeurIPS})

@String(TOG   = {ACM TOG})

@String(ICLR  = {ICLR})

@inproceedings{song2021ddim,
  title     = {Denoising Diffusion Implicit Models},
  author    = {Jiaming Song and Chenlin Meng and Stefano Ermon},
  booktitle = ICLR,
  year      = {2021},
}

@inproceedings{zhang2023adding,
  title={Adding Conditional Control to Text-to-Image Diffusion Models}, 
  author={Lvmin Zhang and Anyi Rao and Maneesh Agrawala},
  booktitle=ICCV,
  year={2023},
}

@article{mvedit2024,
    title={Generic 3D Diffusion Adapter Using Controlled Multi-View Editing},
    author={Hansheng Chen and Ruoxi Shi and Yulin Liu and Bokui Shen and Jiayuan Gu and Gordon Wetzstein and Hao Su and Leonidas Guibas},
    year={2024},
    journal={arXiv:2403.12032},
}

@inproceedings{huberman2024edit,
  title={An edit friendly {DDPM} noise space: Inversion and manipulations},
  author={Huberman-Spiegelglas, Inbar and Kulikov, Vladimir and Michaeli, Tomer},
  booktitle=CVPR,
  year={2024}
}

@inproceedings{chen2024dge,
  title={DGE: Direct Gaussian 3D Editing by Consistent Multi-view Editing},
  author={Minghao Chen and Iro Laina and Andrea Vedaldi},
  booktitle=ECCV,
  year={2024}
}

@inproceedings{lee2025editsplat,
      title={EditSplat: Multi-View Fusion and Attention-Guided Optimization for View-Consistent 3D Scene Editing with 3D Gaussian Splatting},
      booktitle=CVPR,
      author={Dong In Lee and Hyeongcheol Park and Jiyoung Seo and Eunbyung Park and Hyunje Park and Ha Dam Baek and Sangheon Shin and Sangmin Kim and Sangpil Kim},
      year={2025}
}

@inproceedings{LaRa,
         author = {Anpei Chen and Haofei Xu and Stefano Esposito and Siyu Tang and Andreas Geiger},
         title = {LaRa: Efficient Large-Baseline Radiance Fields},
         booktitle = ECCV,
         year = {2024},
        }

@article{hong2023lrm,
  title={Lrm: Large reconstruction model for single image to 3d},
  author={Hong, Yicong and Zhang, Kai and Gu, Jiuxiang and Bi, Sai and Zhou, Yang and Liu, Difan and Liu, Feng and Sunkavalli, Kalyan and Bui, Trung and Tan, Hao},
  journal={arXiv preprint arXiv:2311.04400},
  year={2023}
}

@article{objaverseXL,
  title={Objaverse-XL: A Universe of 10M+ 3D Objects},
  author={Matt Deitke and Ruoshi Liu and Matthew Wallingford and Huong Ngo and
          Oscar Michel and Aditya Kusupati and Alan Fan and Christian Laforte and
          Vikram Voleti and Samir Yitzhak Gadre and Eli VanderBilt and
          Aniruddha Kembhavi and Carl Vondrick and Georgia Gkioxari and
          Kiana Ehsani and Ludwig Schmidt and Ali Farhadi},
  journal={arXiv preprint arXiv:2307.05663},
  year={2023}
}

@inproceedings{liu2023zero1to3,
      title={Zero-1-to-3: Zero-shot One Image to 3D Object}, 
      author={Ruoshi Liu and Rundi Wu and Basile Van Hoorick and Pavel Tokmakov and Sergey Zakharov and Carl Vondrick},
      year={2023},
      booktitle=ICCV
}

@inproceedings{xiang2024structured,
    title   = {Structured 3D Latents for Scalable and Versatile 3D Generation},
    author  = {Xiang, Jianfeng and Lv, Zelong and Xu, Sicheng and Deng, Yu and Wang, Ruicheng and Zhang, Bowen and Chen, Dong and Tong, Xin and Yang, Jiaolong},
    year={2025},
    booktitle=CVPR
}

@article{hui2022neural,
  title     = {Neural Wavelet-Domain Diffusion for 3D Shape Generation},
  author    = {Ka-Hei Hui and Ruihui Li and Jingyu Hu and Chi-Wing Fu},
  journal = TOG,
  year      = {2022}
}

@inproceedings{liu2023one2345++,
  title={One-2-3-45++: Fast Single Image to 3D Objects with Consistent Multi-View Generation and 3D Diffusion},
  author={Minghua Liu and Ruoxi Shi and Linghao Chen and Zhuoyang Zhang and Chao Xu and Xinyue Wei and Hansheng Chen and Chong Zeng and Jiayuan Gu and Hao Su},
  year={2024},
booktitle = CVPR
}

@inproceedings{zhuang2024gtr,
      title={GTR: Improving Large 3D Reconstruction Models through Geometry and Texture Refinement}, 
      author={Peiye Zhuang and Songfang Han and Chaoyang Wang and Aliaksandr Siarohin and Jiaxu Zou and Michael Vasilkovsky and Vladislav Shakhrai and Sergey Korolev and Sergey Tulyakov and Hsin-Ying Lee},
      booktitle=ICLR,
      year={2025} 
}

@inproceedings{wu2024gaussctrl,
    author = {Wu, Jing and Bian, Jia-Wang and Li, Xinghui and Wang, Guangrun and Reid, Ian and Torr, Philip and Prisacariu, Victor},
    title = {{GaussCtrl: Multi-View Consistent Text-Driven 3D Gaussian Splatting Editing}},
    booktitle = ECCV,
    year = {2024},
}

@inproceedings{sella2023vox,
 title={Vox-E: Text-guided Voxel Editing of 3D Objects},
 author={Sella, Etai and Fiebelman, Gal and Hedman, Peter and Averbuch-Elor, Hadar},
 booktitle=ICCV,
 year={2023}
}

@inproceedings{hertz2022prompt,
  title = {Prompt-to-Prompt Image Editing with Cross Attention Control},
  author = {Hertz, Amir and Mokady, Ron and Tenenbaum, Jay and Aberman, Kfir and Pritch, Yael and Cohen-Or, Daniel},
  booktitle=ICLR,
  year = {2024},
}

@inproceedings{lin2025diffsplat,
  title={DiffSplat: Repurposing Image Diffusion Models for Scalable 3D Gaussian Splat Generation},
  author={Lin, Chenguo and Pan, Panwang and Yang, Bangbang and Li, Zeming and Mu, Yadong},
  booktitle=ICLR,
  year={2025}
}

@inproceedings{zhang2024gaussiancube,
  title={GaussianCube: Structuring Gaussian Splatting using Optimal Transport for 3D Generative Modeling},
  author={Zhang, Bowen and Cheng, Yiji and Yang, Jiaolong and Wang, Chunyu and Zhao, Feng and Tang, Yansong and Chen, Dong and Guo, Baining},
  booktitle=NIPS,
  year={2024}
}

@article{Jun2023ShapEGC,
  title={Shap-E: Generating Conditional 3D Implicit Functions},
  author={Heewoo Jun and Alex Nichol},
  journal={arXiv:2305.02463},
  year={2023},
}

@inproceedings{Koo2024PDS,
  title     = {Posterior Distillation Sampling},
  author    = {Koo, Juil and Park, Chanho and Sung, Minhyuk},
  booktitle = CVPR,
  year      = {2024}
}

@inproceedings{chen2023gaussianeditor,
    title={GaussianEditor: Swift and Controllable 3D Editing with Gaussian Splatting},
    author={Yiwen Chen and Zilong Chen and Chi Zhang and Feng Wang and Xiaofeng Yang and Yikai Wang and Zhongang Cai and Lei Yang and Huaping Liu and Guosheng Lin},
    year={2023},
    booktitle = CVPR,
}

@inproceedings{Chen2024proedit,
    title={{ProEdit}: Simple Progression is All You Need for High-Quality {3D} Scene Editing},
    author={Chen, Jun-Kun and Wang, Yu-Xiong},
    booktitle=NIPS,
    year={2024}
}

@article{li2023focaldreamer,
    title={FocalDreamer: Text-driven 3D Editing via Focal-fusion Assembly}, 
    author={Yuhan Li and Yishun Dou and Yue Shi and Yu Lei and Xuanhong Chen and Yi Zhang and Peng Zhou and Bingbing Ni},
    journal={ArXiv},
    year={2023}
}

@inproceedings{erkoc2024preditor3d,
    title={PrEditor3D: Fast and Precise 3D Shape Editing}, 
    author={Ziya Erkoç and Can Gümeli and Chaoyang Wang and Matthias Nießner and Angela Dai and Peter Wonka and Hsin-Ying Lee and Peiye Zhuang},
    year={2024},
    booktitle=CVPR
}

@inproceedings{ho2020denoising,
  title={Denoising Diffusion Probabilistic Models},
  author={Jonathan Ho and Ajay Jain and Pieter Abbeel},
  year={2020},
  booktitle=NIPS
}

@inproceedings{rombach2021highresolution,
title={High-Resolution Image Synthesis with Latent Diffusion Models}, 
author={Robin Rombach and Andreas Blattmann and Dominik Lorenz and Patrick Esser and Björn Ommer},
year={2021},
booktitle=CVPR
}

@inproceedings{wu2023gpteval3d,
   author = {Tong Wu and Guandao Yang and Zhibing Li and Kai Zhang and 
      Ziwei Liu and Leonidas Guibas and Dahua Lin and Gordon Wetzstein},
   title = {GPT-4V(ision) is a Human-Aligned Evaluator for Text-to-3D Generation},
   booktitle = CVPR,
   year = {2024}
}

@inproceedings{Haque2023instructnerf,
    author = {Haque, Ayaan and Tancik, Matthew and Efros, Alexei and Holynski, Aleksander and Kanazawa, Angjoo},
    title = {Instruct-NeRF2NeRF: Editing 3D Scenes with Instructions},
    booktitle = CVPR,
    year = {2023}
}

@inproceedings{si2023freeu,
  title={FreeU: Free Lunch in Diffusion U-Net},
  author={Si, Chenyang and Huang, Ziqi and Jiang, Yuming and Liu, Ziwei},
  booktitle=CVPR,
  year={2024}
}

@inproceedings{mildenhall2020nerf,
 title={NeRF: Representing Scenes as Neural Radiance Fields for View Synthesis},
 author={Ben Mildenhall and Pratul P. Srinivasan and Matthew Tancik and Jonathan T. Barron and Ravi Ramamoorthi and Ren Ng},
 year={2020},
 booktitle=ECCV
}

@Article{kerbl3Dgaussians,
      author       = {Kerbl, Bernhard and Kopanas, Georgios and Leimk{\"u}hler, Thomas and Drettakis, George},
      title        = {3D Gaussian Splatting for Real-Time Radiance Field Rendering},
      journal      = TOG,
      year         = {2023}
}

@inproceedings{Shue2023TriDiff,
  author={Shue, J. Ryan and Chan, Eric Ryan and Po, Ryan and Ankner, Zachary and Wu, Jiajun and Wetzstein, Gordon},
  booktitle=CVPR, 
  title={3D Neural Field Generation Using Triplane Diffusion}, 
  year={2023}
}

@inproceedings{poole2022dreamfusion,
  author = {Poole, Ben and Jain, Ajay and Barron, Jonathan T. and Mildenhall, Ben},
  title = {DreamFusion: Text-to-3D using 2D Diffusion},
  booktitle=ICLR,
  year = {2023}
}

@article{zhuang2023dreameditor,
  title={DreamEditor: Text-Driven 3D Scene Editing with Neural Fields},
  author={Zhuang, Jingyu and Wang, Chen and Liu, Lingjie and Lin, Liang and Li, Guanbin},
  journal= TOG,
  year={2023}
}

@inproceedings{zeng2022lion,
    title={LION: Latent Point Diffusion Models for 3D Shape Generation},
    author={Xiaohui Zeng and Arash Vahdat and Francis Williams and Zan Gojcic and Or Litany and Sanja Fidler and Karsten Kreis},
    booktitle=NIPS,
    year={2022}
}

@inproceedings{Zhou2025diffgs,
    title={DiffGS: Functional Gaussian Splatting Diffusion},
    author={Zhou, Junsheng and Zhang, Weiqi and Liu, Yu-Shen},
    booktitle=NIPS,
    year={2024}
}

@article{nichol2022pointe,
      title={Point-E: A System for Generating 3D Point Clouds from Complex Prompts}, 
      author={Alex Nichol and Heewoo Jun and Prafulla Dhariwal and Pamela Mishkin and Mark Chen},
      year={2022},
      journal={arXiv:2212.08751},
}

@inproceedings{anciukevicius2022renderdiffusion,
	title        = {RenderDiffusion: Image Diffusion for {3D} Reconstruction, Inpainting and Generation},
	author       = {Titas Anciukevicius and Zexiang Xu and Matthew Fisher and Paul Henderson and Hakan Bilen and Mitra, Niloy J. and Paul Guerrero},
	year         = 2022,
	booktitle      = CVPR
}

@InProceedings{yang2024prometheus,
  title={Prometheus: 3D-Aware Latent Diffusion Models for Feed-Forward Text-to-3D Scene Generation},
  author={Yuanbo, Yang and Jiahao, Shao and Xinyang, Li and Yujun, Shen and Andreas, Geiger and Yiyi, Liao},
  booktitle=CVPR,
  year={2025}
}

@inproceedings{chen2024shap,
  title={SHAP-EDITOR: Instruction-guided Latent 3D Editing in Seconds},
  author={Chen, Minghao and Xie, Junyu and Laina, Iro and Vedaldi, Andrea},
  booktitle=CVPR,
  year={2024}
}

@InProceedings{brooks2023instructpix2pix,
    author     = {Brooks, Tim and Holynski, Aleksander and Efros, Alexei A.},
    title      = {InstructPix2Pix: Learning to Follow Image Editing Instructions},
    booktitle  = CVPR,
    year       = {2023},
}

@InProceedings{liu2023grounding,
  title={Grounding dino: Marrying dino with grounded pre-training for open-set object detection},
  author={Liu, Shilong and Zeng, Zhaoyang and Ren, Tianhe and Li, Feng and Zhang, Hao and Yang, Jie and Li, Chunyuan and Yang, Jianwei and Su, Hang and Zhu, Jun and others},
  booktitle=ECCV,
  year={2024}
}

@inproceedings{Deitke2023objaverse,
  title={Objaverse: A Universe of Annotated 3D Objects},
  author={Matt Deitke and Dustin Schwenk and Jordi Salvador and Luca Weihs and
          Oscar Michel and Eli VanderBilt and Ludwig Schmidt and
          Kiana Ehsani and Aniruddha Kembhavi and Ali Farhadi},
  booktitle=CVPR,
  year={2023}
}

@inproceedings{Downs2022GSO,
author = {Downs, Laura and Francis, Anthony and Koenig, Nate and Kinman, Brandon and Hickman, Ryan and Reymann, Krista and McHugh, Thomas B. and Vanhoucke, Vincent},
title = {Google Scanned Objects: A High-Quality Dataset of 3D Scanned Household Items},
year = {2022},
booktitle = ICRA,
}

@article{ravi2024sam2,
  title={SAM 2: Segment Anything in Images and Videos},
  author={Ravi, Nikhila and Gabeur, Valentin and Hu, Yuan-Ting and Hu, Ronghang and Ryali, Chaitanya and Ma, Tengyu and Khedr, Haitham and R{\"a}dle, Roman and Rolland, Chloe and Gustafson, Laura and Mintun, Eric and Pan, Junting and Alwala, Kalyan Vasudev and Carion, Nicolas and Wu, Chao-Yuan and Girshick, Ross and Doll{\'a}r, Piotr and Feichtenhofer, Christoph},
  journal={arXiv:2408.00714},
  year={2024}
}

@misc{igs2gs,
         author = {Vachha, Cyrus and Haque, Ayaan},
         title = {Instruct-GS2GS: Editing 3D Gaussian Splats with Instructions},
         year = {2024},
         url = {https://instruct-gs2gs.github.io/}
        }

@inproceedings{huang2025edit360,
  title={Edit360: 2D Image Edits to 3D Assets from Any Angle},
  author={Huang, Junchao and Hu, Xinting and Shi, Shaoshuai and Tian, Zhuotao and Jiang, Li},
  booktitle=ICCV,
  year={2025}
}

@article{li2025cmd,
  title={CMD: Controllable Multiview Diffusion for 3D Editing and Progressive Generation},
  author={Li, Peng and Ma, Suizhi and Chen, Jialiang and Liu, Yuan and Zhang, Chongyi and Xue, Wei and Luo, Wenhan and Sheffer, Alla and Wang, Wenping and Guo, Yike},
  journal={arXiv preprint arXiv:2505.07003},
  year={2025}
}

@inproceedings{trellis,
    title   = {Structured 3D Latents for Scalable and Versatile 3D Generation},
    author  = {Xiang, Jianfeng and Lv, Zelong and Xu, Sicheng and Deng, Yu and Wang, Ruicheng and 
               Zhang, Bowen and Chen, Dong and Tong, Xin and Yang, Jiaolong},
    booktitle = CVPR,
    year    = {2025}
}

@inproceedings{wen2025intergsedit,
  title     = {InterGSEdit: Interactive 3D Gaussian Splatting Editing with 3D Geometry-Consistent Attention Prior},
  author    = {Wen, Minghao and Wu, Shengjie and Wang, Kangkan and Liang, Dong and others},
  booktitle = ICCV,
  year      = {2025}
}

@inproceedings{wu2024freeinit,
  title     = {FreeInit: Bridging Initialization Gap in Video Diffusion Models},
  author    = {Wu, Tianxing and Si, Chenyang and Jiang, Yuming and Huang, Ziqi and Liu, Ziwei},
  booktitle = ECCV,
  year      = {2024},
}
